\documentclass[pra,aps,floats,floatfix,twocolumn,showpacs,longbibliography,superscriptaddress,notitlepage,preprintnumbers,10pt]{revtex4-1}
\usepackage[USenglish]{babel}
\usepackage{amsmath,amssymb,amsfonts}
\usepackage{graphicx}
\usepackage{dcolumn}
\usepackage{bm}
\usepackage[normalem]{ulem} 
\usepackage[colorlinks=true,linkcolor=blue,citecolor=red]{hyperref}
\usepackage[T1]{fontenc} 
\usepackage{lmodern}
\usepackage[usenames,dvipsnames]{color}

\newcommand{\wymiar}{0.45\textwidth}

\newcommand{\wymiarczt}{0.375 \textwidth}

\begin{document}
      \title{Dynamical leakage of Majorana mode into side-attached quantum dot}

     \author{J.\ Bara\'nski}
     \email[e-mail: ]{j.baranski@law.mil.pl}
     \affiliation{Military University of Aviation, ul. Dywizjonu 303, 08-521 D\k{e}blin, Poland}
     
     \author{M.\ Bara\'nska}
     \affiliation{Military University of Aviation, ul. Dywizjonu 303, 08-521 D\k{e}blin, Poland}
     
      \author{T.\ Zienkiewicz }
      \affiliation{Military University of Aviation, ul. Dywizjonu 303, 08-521 D\k{e}blin, Poland}
     
      \author{R.\ Taranko}
     \affiliation{Institute of Physics, 
	     M.\ Curie-Sk\l odowska University, 
             20-031 Lublin, Poland}

      \author{T.\ Doma\'nski }
     \email[e-mail: ]{doman@kft.umcs.lublin.pl}
     \affiliation{Institute of Physics, 
	     M.\ Curie-Sk\l odowska University, 
             20-031 Lublin, Poland}

      \date{\today}
 

\begin{abstract}
We study a hybrid structure, comprising the single-level quantum dot attached to the topological superconducting nanowire,  inspecting dynamical transfer of the Majorana quasiparticle onto normal region. Motivated by the recent experimental realization of such heterostructure and its investigation under the stationary conditions 
[L. Schneider {\it et al.}, \href{https://doi.org/10.1038/s41467-020-18540-3}
 {Nature Communications {\bf 11}, 4707 (2020)}] 
where the quantum dot energy level can be tuned by gate potential we examine how much time is needed for the Majorana mode to leak into the normal region. We estimate, that  
for typical hybrid structures this dynamical process would take about 20 nanoseconds. We propose a feasible empirical protocol for its detection by means of the time-resolved Andreev tunneling spectroscopy.
\end{abstract}

\maketitle

\section{Motivation}

Topological superconductors, hosting the Majorana boundary modes, are currently of great interests both for basic science \cite{Aguado.2017,Lutchyn.2018,Kouwenhoven-2020} and potential applications \cite{Aguado-2020,Liu-2016}. Topological protection and nonabelian character make them promising candidates for realization of stable qubits \cite{Schrade-2018} and quantum computations \cite{Karzig-2017}. Signatures of their fractional statistics can tested, for instance, by a sequence of the charge-transfer operations using the quantum dots attached to topological superconductor \cite{Flensberg-2011,Steiner-2020}. Such dynamical transfer of the charge between the quantum dot and  topological superconductors might enable nonabelian operations on the Majorana bound states  \cite{Souto-2020,Thorwart-2020}. 

Development of time-resolved spectroscopies with their resolution down to sub-picosecond regime allows to probe the physical structures in response to abrupt change of the model parameter(s) or other nonequilibrium conditions \cite{Tuovinen_2019}. Such dynamics has been recently investigated in topological phases by a number of groups, considering the fermionic \cite{Mazza-2015,Wang-2017,Yang-2018} and bosonic systems \cite{Maffei-2018}. In particular, there has explored dynamical teleportation due to nonlocality of the zero-energy boundary modes \cite{Li-2020} and it has been shown, that dynamical techniques based on the noise measurements could unambiguously identify the true Majorana quasiparticles \cite{Jonckheere-2019,Manousakis-2020,Radgohar-2020,Civelli_Simon-2020}. Time-dependent measurements have been also proposed to detect topological invariants of the higher order topological superconductor, harboring the corner states \cite{Mizoguchi-2020}. 

Hybrid structures consisting of the topological superconducting nanowires side-attached to the quantum dots allow for a tunable control of the zero-energy bound states \cite{Deng-2016,Deng-2018,Wiesendanger-2020}. Their variations with respect to the gate potentials or magnetic field could discriminate the true Majorana quasiparticles from the trivial Andreev bound states, appearing accidentally at zero energy \cite{Sarma-2020}. More complex magnetic-superconducting heterostructures enable coexistence of the localized and chiral modes \cite{Morr-2020}.

\begin{figure}
	\includegraphics[width=\wymiarczt]{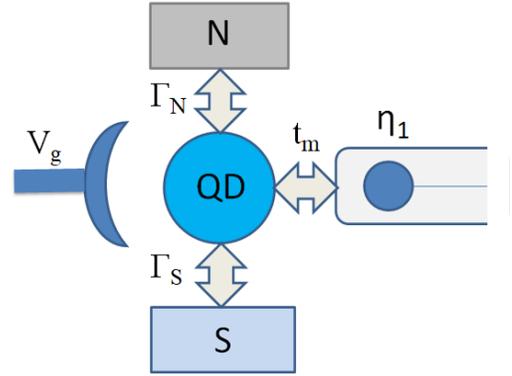}
	\caption{Schematic representation of considered system comprising quantum dot (QD) deposited between metallic (N) and superconducting (S) electrode and side coupled to one end of monoatomic chain hosting Majorana edge state represented by $\eta_1$. Additional electrode stands for gate voltage ($V_g$) tuning of quantum dot's energy level.} 
	\label{scheme}
\end{figure}

Here we analyze a dynamical transfer of the Majorana mode from the superconducting nanowire to the quantum dot (see  Fig.\ \ref{scheme}). We consider such evolution driven by their sudden coupling or abrupt change of the QD energy level imposed by applying external gate potential. We examine how much time is needed for the Majorana mode to leak into the quantum dot region. We also 
show, that time-resolved process provide valuable insight into the effective quasiparticles of topological superconductors.

The paper is organized as follows. In Sec. \ref{sec:model} we introduce the microscopic model and outline the procedure for studying the time-dependent evolution. Next, in Sec. \ref{sec:leakage} we present the time-dependent signatures of the Majorana quasiparticle leaking on the side-attached quantum dot. In Sec. \ref{sec:quantitative_evaluation} we provide realistic estimations of the characteristic time- and energy-scales which could be practically verified experimentally.  Finally, Sec.~\ref{sec:summary} summarizes the main results.

\section{Formulation of the problem}
\label{sec:model}

We study dynamical properties of the heterostructure comprising the quantum dot deposited on s-wave superconductor and weakly coupled to another metallic lead  that can be thought as STM tip. The quantum dot is additionally coupled to the topological superconducting nanowire, hosting the Majorana end modes  (Fig.~\ref{scheme}). Our considerations aim to determine the characteristic time needed for development of the Majorana features transmitted onto the QD region. In particular, we shall investigate the quantum evolution (i) driven by abrupt formation of the hybrid structure, inspecting the zero-energy features appearing for various positions of the quantum dot energy level, and b) study how long it takes to qualitatively transform the existing Majorana signatures after a sudden change of the QD level.

It has been predicted \cite{Vernek-2014} that in QD attached to the topological superconductor the zero-energy quasiparticle would be transmitted to the normal region. The first empirical evidence for such Majorana leakage has been reported by M.T. Deng {\it et al} \cite{Deng-2016} and recently in STM method by L. Schneider {\it et al} \cite{Wiesendanger-2020}. The quantum dot - topological superconductor hybrid structures have studied by a number of groups (see Ref.\ \cite{DasSarma-2020} for survey). In N-QD-S configurations (Fig.~\ref{scheme}) such leakage process qualitatively depends on the QD level \cite{JBTDAK2017,JBTD2019}. Namely, when the energy level is away from zero, the signatures of Majorana mode  are manifested by the peak appearing in the density of states (DOS), thereby enhancing the zero-bias conductance.
Contrary to such picture, when the QD level happens to be near zero energy, the quantum interference induces the fractional Fano-type  depletion in the QD spectrum \cite{Orellana-2019}. Dynamical changeover from one regime to another has not been analyzed yet, and this is the main purpose of our present study.

\subsection{Microscopic scenario}

Let us start, by formulating the microscopic model and presenting the computational method used for determination of time-dependent quantities. We describe our hybrid structure (Fig.~\ref{scheme}) by the following Hamiltonian
\begin{eqnarray}
\hat{H}=\hat{H}_{QD}+\sum_{\beta} \left( \hat{H}_{\beta}+ \hat{T}_{QD-\beta}\right) +\hat{H}_{MQD} ,
\label{H1}
\end{eqnarray}
where $\hat{H}_{QD}$ refers to the quantum dot (QD), $\hat{H}_{\beta}$ denote the metallic ($\beta=N$) and superconducting ($\beta=S$) leads, whereas $\hat{H}_{MQD}$ stands for the zero-energy Majorana quasiparticles and coupling of one of them to QD. 
The normal lead can be treated as a Fermi sea $\hat{H}_{N}=\sum_{k,\sigma} \xi_{N k} \hat{c}^{\dagger}_{N k\sigma}\hat{c}_{N k\sigma }$, where $\hat{c}^{\dagger}_{N k \sigma}$ $(\hat{c}_{N k \sigma})$ is the creation (annihilation) of $\sigma$ spin electron whose energy $\xi_{N k}=\varepsilon_{Nk}-\mu_{N}$ is measured from the chemical potential $\mu_{N}$. The superconducting electrode is taken in BCS-type form $\hat{H}_{S}=\sum_{k,\sigma} \xi_{S k} \hat{c}^{\dagger}_{S k\sigma}\hat{c}_{S k\sigma }-\sum_{k} \left( \Delta_{S} \hat{c}^{\dagger}_{S k\uparrow} \hat{c}^{\dagger}_{S k\downarrow} + h.c. \right)$, where $\Delta_{S}$ is the isotropic pairing gap. Itinerant electrons are hybridized with the quantum dot through the term
$\hat{T}_{QD-\beta}= \sum_{k, \sigma} \left( V_{\beta  k}  \hat{d}^{\dagger}_{ \sigma} \hat{c}_{\beta k \sigma}+ h.c.\right)$. We introduce the coupling functions $\Gamma_{\beta}=2\pi\sum_{k} \left| V_{\beta  k}\right|^{2} \delta \left( \omega-\xi_{\beta k} \right)$ assuming them to be constant in the low-energy regime.

As regards the quantum dot we represent it by the single level impurity term $\hat{H}_{QD}=\sum_{\sigma} \epsilon_{d} \hat{d}_{\sigma}^{\dagger}\hat{d}_{ \sigma}$, where $\hat{d}^{\dagger}_{\sigma}$ ($\hat{d}_{\sigma}$) is the creation (annihilation) operator of spin $\sigma=\uparrow$, $\downarrow$ electrons. We focus the low energy regime $|\omega| \ll \Delta_S$, when the fermion degrees of freedom outside the pairing gap can be integrated out. Their influence simplifies to the induced pairing $\hat{H}_{S}+\hat{T}_{QD-S} \approx \Delta_d \left( \hat{d}_{ \downarrow} \hat{d}_{ \uparrow} +\mbox{\rm h.c.} \right)$ with the on-dot potential $\Delta_{d}=\Gamma_{S}/2$ \cite{Bauer_2007,JBTD_2013,Yamada2011}. 

The `proximitized' QD is thus effectively modeled by
\begin{eqnarray}
\hat{H}_{QD}\!+\!\hat{H}_{S}\!+\!\hat{T}_{QD-S}\!=\!\sum_{\sigma}\! \epsilon_{d} \hat{d}_{ \sigma}^{\dagger}\hat{d}_{ \sigma}\!+\!\Delta_d \left( \hat{d}_{ \downarrow} \hat{d}_{ \uparrow} \!+\!\mbox{\rm h.c.} \right)\!.
\end{eqnarray}
Restricting our considerations to the low-energy regime (within milielectronvolts around the chemical potential) we describe the topological superconductor  by
\begin{eqnarray}
\hat{H}_{MQD}= i\epsilon_{m}\hat{\eta}_{1}\hat{\eta}_{2}+\lambda(d_{ \uparrow}\hat{\eta}_{1}+ \hat{\eta}_{1}\hat{d}^{\dagger}_{\uparrow}),
\label{MQD1}
\end{eqnarray}
where the selfhermitian operators $\hat{\eta}_1$, $\hat{\eta}_2$ represent the Majorana end-modes and $\epsilon_{m}$ stands for their overlap. The last part appearing in Eq.\ (\ref{MQD1}) accounts for hybridization of the left h.s.\ Majorana  quasiparticle with the quantum dot, where $\lambda$ is the coupling strength. For convenience we recast the Majorana operators by the standard fermion operators defined via $\hat{\eta}_{1}=\frac{1}{\sqrt{2}}(\hat{f}+\hat{f}^{\dagger})$, $\eta_{2}=\frac{-i}{\sqrt{2}}(f-f^{\dagger})$. This transformation implies, that (\ref{MQD1}) can be rewritten as
\begin{eqnarray}
\hat{H}_{MQD}&=&\epsilon_{m}\hat{f}^{\dagger}\hat{f}+t_{m}(\hat{d}_{\uparrow}^{\dagger}-\hat{d}_{ \uparrow})(\hat{f}+\hat{f}^{\dagger}) 
\label{effective_MQD}
\end{eqnarray} 
with a shorthand notation $t_m= \lambda /\sqrt{2}$.

\subsection{Outline of computational method}

Fingerprints of the Majorana mode would be practically observed in the quantum dot region, examining the charge current $I_{\sigma}(V,t)$ induced via N-QD-S junction by the bias voltage $V$. Its differential conductance $G_{\sigma}(V,t)=\frac{d}{dV} I_{\sigma}(V,t)$ can be obtained numerically, using the following expression for the current
\begin{eqnarray}
I_{\sigma}(V,t)= -e \left< \frac{d\hat{N}_{N\sigma}}{dt} \right> 
= -\frac{i e}{\hbar} \langle [\hat{H},\hat{N}_{N\sigma}] \rangle.
\label{i1}
\end{eqnarray}
Since all parts of the Hamiltonian (\ref{H1}), except $\hat{T}_{QD-N}$, commute with the number operator $\hat{N}_{N\sigma}$ therefore the tunneling current reads 
\begin{eqnarray}
I_{\sigma}(V,t) & = & \frac{-ie}{\hbar} \sum_{k} \left( V_{N  k} \langle \hat{d}^{\dagger}_{\sigma}(t) \hat{c}_{N k \sigma}(t) \rangle 
- \mbox{\rm h.c.} \right) \nonumber \\
&=& \frac{2e}{\hbar} \; \Im \left[ \sum_{k} V_{N k} \langle \hat{d}^{\dagger}_{\sigma}(t) \hat{c}_{N k \sigma}(t) \rangle \right] .
\label{i2}
\end{eqnarray}
To simplify the notation from now onwards  we set $\hbar=e=\Gamma_{N}=1$, expressing the energies, currents  and time in units of $\Gamma_{N}$, $e\Gamma_{N}/\hbar$ and $\hbar/\Gamma_{N}$, respectively. Additionally we choose the chemical potential of superconductor as a reference level ($\mu_{S}=0$), when $\mu_{N}=eV$. Since the itinerant electrons degrees of freedom refer solely to the metallic lead, for brevity we also skip the subindex $N$ appearing in  $\hat{c}^{(\dagger)}_{N k \sigma}$, $V_{N k}$, and $\xi_{N k}$. 

Using the standard expression for $\hat{c}_{k \sigma}(t)$ \cite{Taranko-2018}
\begin{eqnarray}
\hat{c}_{k \sigma} (t)=\hat{c}_{k\sigma}(0)e^{-i \xi_{k} t}-iV_{k}\int_{0}^{t}d\tau e^{-i\xi_{k}(t-\tau)}\hat{d}_{\sigma}(\tau)
\label{c(t)}
\end{eqnarray}
and imposing the wide band limit approximation we get
\begin{eqnarray}
I_{\sigma}(t)=2\Im \left[\sum_{k} V_{k} \langle \hat{d}^{\dagger}_{\sigma}(t)\hat{c}_{k \sigma}(0) \rangle - i \frac{\Gamma_{N}}{2}\langle \hat{n}_{\sigma}(t) \rangle \right].
\label{eq:prad}
\end{eqnarray} 
Here $\langle ... \rangle$ denotes the statistically averaged value. In  order to determine the current (\ref{i2}) one needs the correlation function $\langle \hat{d}_{\sigma}^{\dagger}(t) \hat{c}_{k \sigma}(0)\rangle$  and the time-dependent QD occupancy $\langle \hat{n}_{\sigma}(t) \rangle=\langle \hat{d}_{\sigma}^{\dagger}\hat{d}_{\sigma}\rangle$. 
 
To find these quantities we employ the equation of motion approach (EOM) \cite{Taranko-2018} and make use of the 4-th order Runge-Kutta method to solve numerically the appropriate set of differential equations. 
Traditionally such EOM applied to any correlation function generates a sequence of the additional functions, for which the next equations of motion must be constructed, until they are finally closed (or terminated).
For the uncorrelated QD embedded in our heterostructure this procedure mixes the operators $\hat{d}^{(\dagger)}_{\sigma}$  and $\hat{f}^{(\dagger)}$ with the operators of the normal lead electrons. 
They have a general form $\sum_{k}V_{k} \langle \hat{O}(t)\hat{c}_{k \sigma}(0) \rangle$, where $\hat{O}$ stands for one of the following six operators $\hat{d}_{\downarrow}$, $\hat{d}^{\dagger}_{\downarrow}$, $\hat{d}_{\uparrow}$, $\hat{d}^{\dagger}_{\uparrow}$, $\hat{f}$, and $\hat{f}^{\dagger}$.
We thus have to determine 12 momentum-dependent correlation functions $f_{k i}(t)=\langle \hat{O}(t)\hat{c}_{k \sigma}(0) \rangle$ and 9 momentum-independent expectation values $f_{i}(t)=\langle \hat{O}(t)\hat{\tilde{O}}(t) \rangle$ 
 listed explicitly in table \ref{tab2}. Their  differential equations are presented in the appendix [see Eqs (\ref{diff k functions},\ref{diff nok functions})].

\begin{table}[t!]
  \begin{center}
    \begin{tabular}{c|c}
      \textbf{$f_i$} & \textbf{$f_{k i}$} \\
      \hline
 $f_1 = \langle \hat{d}_{\uparrow}^{\dagger}(t) \hat{d}_{\uparrow}(t)\rangle$ & $f_{k 1} = \langle \hat{d}_{\uparrow}^{\dagger}(t) \hat{c}_{k \uparrow}(0)\rangle$  \\  
 $f_2 = \langle \hat{d}_{\downarrow}^{\dagger}(t) \hat{d}_{\downarrow}(t)\rangle$ & $f_{k 2} = \langle \hat{d}_{\uparrow}(t) \hat{c}_{k \downarrow}(0)\rangle$ \\
 $f_3 = \langle\hat{f}^{\dagger}(t) \hat{f}(t)\rangle$ & $f_{k 3} =  \langle d_{\uparrow}(t) \hat{c}_{k \uparrow}(0)\rangle $ \\
 $f_4 = \langle \hat{d}_{\downarrow}(t) \hat{d}_{\uparrow}(t)\rangle$ & $f_{k 4} = \langle d_{\downarrow}^{\dagger}(t) \hat{c}_{k \downarrow}(0)\rangle$ \\
 $f_5 = \langle \hat{d}_{\uparrow}^{\dagger}(t) \hat{f}(t)\rangle$ & $f_{k 5} = \langle \hat{f}^{\dagger}(t) \hat{c}_{k \uparrow}(0)\rangle$ \\
 $f_6 = \langle \hat{d}_{\uparrow}^{\dagger}(t) \hat{f}^{\dagger}(t)\rangle$ & $f_{k 6} = \langle \hat{f}(t) \hat{c}_{k \uparrow}(0)\rangle$ \\
 $f_7 = \langle \hat{d}_{\downarrow}(t) \hat{f}(t)\rangle$ & $f_{k 7} = \langle \hat{f}(t) \hat{c}_{k \downarrow}(0)\rangle$ \\
 $f_8 = \langle \hat{d}_{\downarrow}(t) \hat{f}^{\dagger}(t)\rangle$ & $f_{k 8} = \langle \hat{f}^{\dagger}(t) \hat{c}_{k \downarrow}(0)\rangle$ \\
 $f_9 = \langle \hat{d}_{\uparrow}^{\dagger}(t) \hat{d}_{\downarrow}(t)\rangle$ & $f_{k 9} = \langle \hat{d}_{\downarrow}(t) \hat{c}_{k \downarrow}(0)\rangle$ \\
  & $f_{k 10} = \langle \hat{d}_{\uparrow}^{\dagger}(t) \hat{c}_{k \downarrow}(0)\rangle$ \\
   & $f_{k 11} = \langle\hat{d}_{\downarrow}(t) \hat{c}_{k \uparrow}(0)\rangle$ \\
    & $f_{k 12} = \langle \hat{d}_{\downarrow}^{\dagger}(t) \hat{c}_{k \uparrow}(0)\rangle$ \\
 \end{tabular}
    \caption{The list of 12 momentum-dependent functions $f_{k i}$ and 9 momentum-independent functions $f_i$ constituting a closed set of the equations of motion.}
    \label{tab2}
  \end{center}
\end{table}

Let us note, that in such treatment the parameters (such as the coupling $t_m$  or the energy level $\epsilon_{\sigma}$) can be either static or may depend on time in arbitrary way.

\section{Time-resolved features}
\label{sec:leakage}

The most profound consequence of bringing the quantum dot in contact with topological superconductor is emergence of the zero-energy quasiparticle in the spectrum of QD. This feature (measurable by charge transport through N-QD-S circuit) is a signature of the Majorana mode. Dynamical leakage process provides information about the time required to perform logical operations with use of the Majorana quasiparticles. 

\subsection{Dynamical Majorana leakage}
\label{sec:transient}

Let us estimate the time required for transferring the zero-energy mode onto QD region, after abruptly coupling the quantum dot to topological chain. Until $t=0$ all parts of our setup (Fig.\ \ref{scheme}) are assumed to be completely disconnected and the quantum dot unoccupied. Thus our initial conditions are $\langle \hat{d}^{\dagger}_{\sigma}(0)\hat{d}_{\sigma}(0)\rangle =0$, $\langle \hat{d}_{\downarrow}(0)\hat{d}_{\uparrow}(0) \rangle=0$, $\langle \hat{c}^{\dagger}_{k \sigma}(0)\hat{c}_{k \sigma}(0) \rangle=f_{FD}(\varepsilon_k-eV)$ and the mixed terms $\langle \hat{O}(0) \hat{c}_{k \sigma}(0)\rangle=0$. At $t=0^{+}$ the QD is abruptly connected to external reservoirs by the couplings $\Gamma_{\beta}$. To ensure the subgap quasiparticle states to be well separated we impose the asymmetric couplings  ($\Gamma_S=3\Gamma_N$). The superonducting proximity leads to gradual buildup of the in-gap Andreev bound states. We noticed that these states reach their equilibrium positions and amplitudes after time $t \simeq \hbar/\Gamma_{N}$ \cite{Taranko-2018}.

\begin{figure}[t!]
	\includegraphics[width=\wymiar]{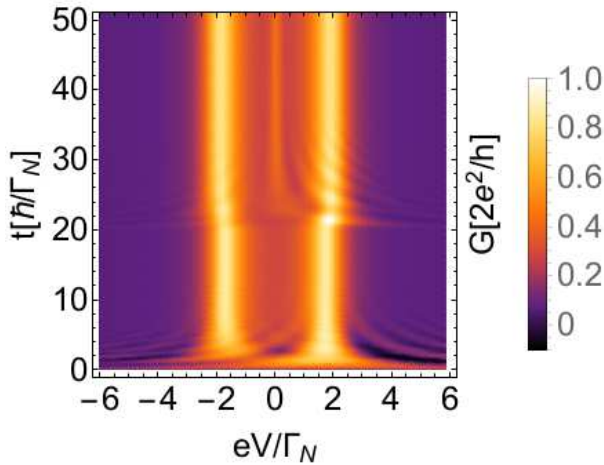}
	\caption{The time-dependent differential conductance as a function of bias $V$ (horizontal axis) and time $t$ (vertical axis) obtained for $\epsilon_{\uparrow}=\epsilon_{\downarrow}=\Gamma_{N}$. At $t=0$ the quantum dot is abruptly connected to the external leads (imposing $\Gamma_S=3\Gamma_N$) when the Andreev bound states (ABS) begin to emerge. Once the position and amplitude of ABS are established at $t=20\hbar/\Gamma_{N}$ the quantum dot is additionally connected to the topological superconductor, assuming $t_{m}=0.5\Gamma_{N}$. From this moment onward the zero-energy peak gradually emerges, signalizing the Majorana leakage into the quantum dot region. For the chosen set of model parameters this zero-energy feature stabilizes its shape after approximately $15-20$ units of time, i.e.\ at $t\simeq40\hbar/\Gamma_{N}$.} 
	\label{fig_evolution}
\end{figure}

Safely later after the Andreev states are stabilized we abruptly connect the quantum dot to topological superconductor. For computations we assume $t_m = 0.5 \Gamma_N$, but more detailed discussion concerning influence of $t_m$ on the time required for the MZM to leak onto QD region is given in Sec.~\ref{sec:quant_estim}. Starting from  $t=20\hbar/\Gamma_{N}$ the Majorana mode gradually leaks to the QD region, as manifested by enhancement of the zero bias conductance (Fig.~\ref{fig_evolution}). We can notice that its amplitude  establishes within the time interval $\Delta t \sim 15-20 \hbar /\Gamma_N$. This result brings us the needed information on the characteristic time of the Majorana leakage. We need to keep in mind, however, that the coupling of QD to metallic reservoir $\Gamma_N$ in different experimental realizations can take various values. To estimate the order of magnitude of the leakage time (in nanoseconds), in Sec.~\ref{sec:evaluation} we compare the qualitative results with typical energy scales used in experiments on in-gap states and hybrids comprising quantum dots and topological superconducting chains.

\subsection{Quench-driven dynamics}
\label{sec:quench}

The transient evolution discussed in Sec.~\ref{sec:transient} allowed us to estimate the time required for the MZM leakage. Abrupt coupling of the QD to topological superconductor would be however hardly feasible in practice. More realistic scenario can rely on employing the gate voltage potential to vary the QD's energy level in controlled manner. Depending on the specific value of $\epsilon_{\sigma}$, the Majorana mode should be evidenced either by the interferometric depletion (when $\epsilon_{\sigma} \simeq 0$) or constructive enhancement (for $|\epsilon_{\sigma}|\gg t_m$ ) of the zero-bias tunneling conductance \cite{JBTDAK2017,JBTD2019}.
To provide some experimentally verifiable result we propose to test how long it takes to transform  the ditch into the peak feature. Time needed for such changeover can be subsequently confronted with the timescale of the MZM leakage driven by abrupt coupling of the QD to the topological chain.

\begin{figure}[t!]
	\includegraphics[width=\wymiar]{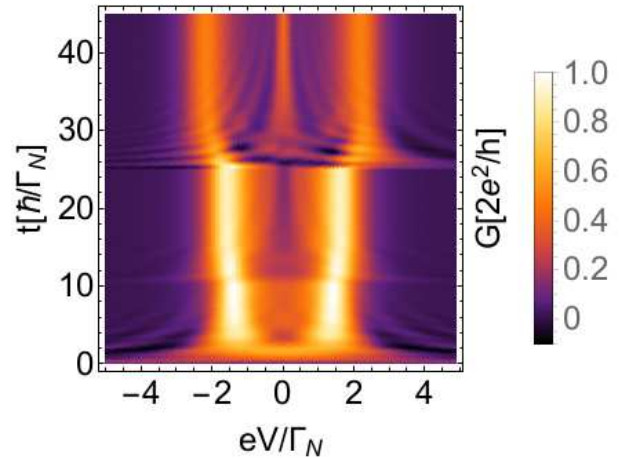}
	\caption{Time-resolved differential conductance obtained for $\Gamma_S=3\Gamma_N$, $t_m=0.5\Gamma_N \theta(t-10\frac{\hbar}{\Gamma_N})$ and the QD energy level $\epsilon_{\sigma}=1.5\Gamma_N \theta(t-25\frac{\hbar}{\Gamma_N})$, where $\theta$ is step function.
	Similarly as in Fig.\ \ref{fig_evolution} we initially couple the QD to the external N/S leads, imposing the energy level $\epsilon_{\sigma}=0$. After $t=10 \hbar/\Gamma_N$ the quantum dot is connected to topological superconductor, inducing the destructive interference ditch in the zero-bias conductance. At $t=25\hbar/\Gamma_N$) we subsequently lift the QD energy level to $\epsilon_{\uparrow}=\epsilon_{\downarrow}=\Gamma_N$ and from this moment onward the ditch transforms into the zero-bias peak. This feature establishes after approximately $15 \hbar/\Gamma_N$,  i.e. at $t\simeq 40\hbar/\Gamma_{N}$.} 
	\label{fig2}
\end{figure}

For this purpose we consider the following three step procedure.  
(i) As previously, we start by forming the N-QD-S circuit with the initial energy level $\epsilon_{\sigma}=0$ and let the Andreev bound states to establish. (ii) Once the differential conductance saturates at its static value (at $t\simeq 10\hbar/\Gamma_{N}$ fluctuations become almost negligible) we couple the quantum dot to the topological chain. After certain amount of time there appears the interference ditch in the zero-bias conductance. (iii) When `the dust is settled' we abruptly lift the QD energy level to $\epsilon_{\sigma}=1.5\Gamma_N$ by applying the gate voltage. In this way the destructive interference would be gradually replaced by the conventional MZM leakage regime. 

We have estimated that this transition takes approximately $\Delta t = 15 [\frac{\hbar}{\Gamma_N}]$. Such timescale to transform one Majorana feature to another is comparable to the time interval needed for emergence of the zero-energy peak after abrupt coupling of the QD to topological superconductor (see Fig.\ \ref{fig_evolution}). Observation of such dynamical changeover could thus indirectly probe the MZM leakage time itself. In Sec.\ \ref{sec:quantitative_evaluation} we provide quantitative evaluation of this characteristic time, having in mind the typical energy scales in experiments using with various quantum dots coupled to superconductors and/or topological superconductors. 

\section{Quantitative evaluations}
\label{sec:quantitative_evaluation}

To deliver reliable information on the timescale in some tangible  units one has to take into account the specific energy scales for the experimentally achievable setups in analysis of the Andreev/Majorana bound states. Let us consider a few realistic examples.

\subsection{Typical energy and time scales}
\label{sec:evaluation}

The energy gap of conventional ($s$-wave) superconductors, which are often used in experiments with quantum dots varies from a few tens to hundreds of microelectronvolts. For instance, vanadium electrode used in Ref.\  \cite{Lee2013} was characterized by $\Delta_{S} \simeq 0.55$ meV. The energy gap of titanium electrode used by Deacon {\it et al.} \cite{Deacon2010} was about $152$ $\mu$eV. In the present context 
more useful would be the proximity induced on-dot pairing gap, which is roughly equal to $\Gamma_{S}$ \cite{JBTD_2013}. Its value in experiment performed by Sch\"onenberger {\it et al.} \cite{BGR2019} varied from $10$ $\mu$eV to $165$ $\mu$eV, whereas in the setup of Deng {\it et al.} \cite{Deng-2016} using Ti/Al its magnitude was $250$ $\mu$eV. 

To observe well pronounced subgap states the hybridization with metallic electrode  $\Gamma_{N}$ (which controls the inverse life time) should be considerably smaller than both $\Gamma_{S}$ and $\Delta_{S}$. For this reason we have enforced $\Gamma_{N}\ll\Gamma_{S}$, otherwise the in-gap states would overlap with each other. In numerous experiments devoted to investigations of the in-gap bound states, such coupling to metallic electrode $\Gamma_{N}$ was kept about ten or hundred times smaller than $\Gamma_{S}$. For example Ref.\ \cite{Lee2013} reported the experimental value $\Gamma_N \simeq 50$ $\mu$eV. In our present approach (where $\Gamma_N$ is used as energy unit) we thus assume the following realistic value $\Gamma_N \simeq 5 - 50$ $\mu$eV, implying the time unit $\frac{\hbar}{\Gamma_N} \simeq 0.125 - 1.25$ nanoseconds. Taking such quantities into account, the time of MZM leakage into the quantum dot region $\simeq 15 [\frac{\hbar}{\Gamma_N}]$ (estimated for $t_m=0.5\Gamma_N$) would be approximately $2 - 20$ ns. 
Transport measurements have temporal resolution in sub-picosecond regime \cite{Tuovinen_2019}, so this dynamical process should be observable in real time.

One should note that the coupling $t_m$ between QD and topological superconducting chain is expressed in terms of $\Gamma_N$. 
To reconcile the specific influence of $t_m$ on the Majorana leakage time, we shall briefly analyze in Sec.\ \ref{sec:quant_estim} a few representative values. As concerns a quantitative effect of the hybridization $\Gamma_S$ we have checked, that its influence on the Majorana mode leakage time onto the QD region is rather negligible.

\subsection{Influence of $t_{m}$}
\label{sec:quant_estim}

\begin{figure}
	\includegraphics[width=\wymiar]{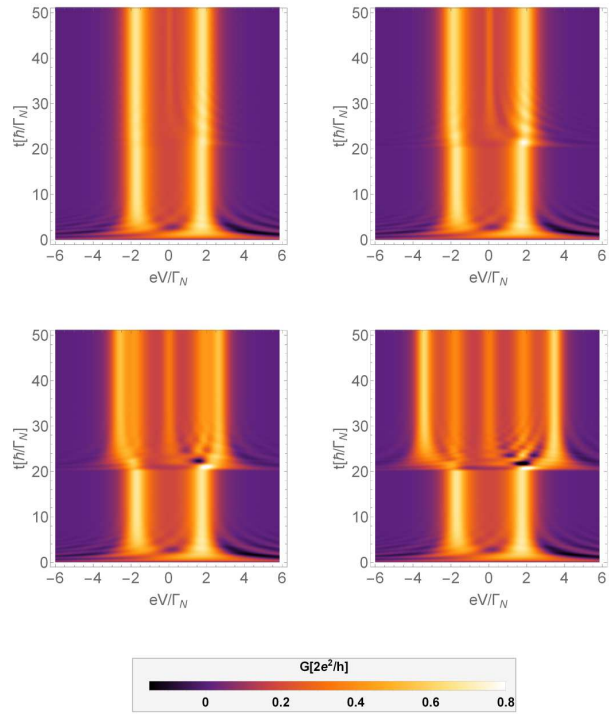}
	\caption{Evolution of the differential conductance obtained for setup shown in Fig.\ \ref{scheme}, assuming the couplings $t_m = 0.25\Gamma_{N}$ (upper left), $0.5\Gamma_{N}$ (upper right), $1\Gamma_{N}$ (lower left), $1.5\Gamma_{N}$ (lower right). The quantum dot is abruptly hybridized with the Majona end-mode at $t=20\hbar/\Gamma_{N}$. We used the model parameters $\Gamma_{S}=3\Gamma_{N}$, $\epsilon_{\sigma}=\Gamma_N$.}
	\label{GridTemp2}
\end{figure}

\begin{figure*}
	\includegraphics[width=0.95\textwidth]{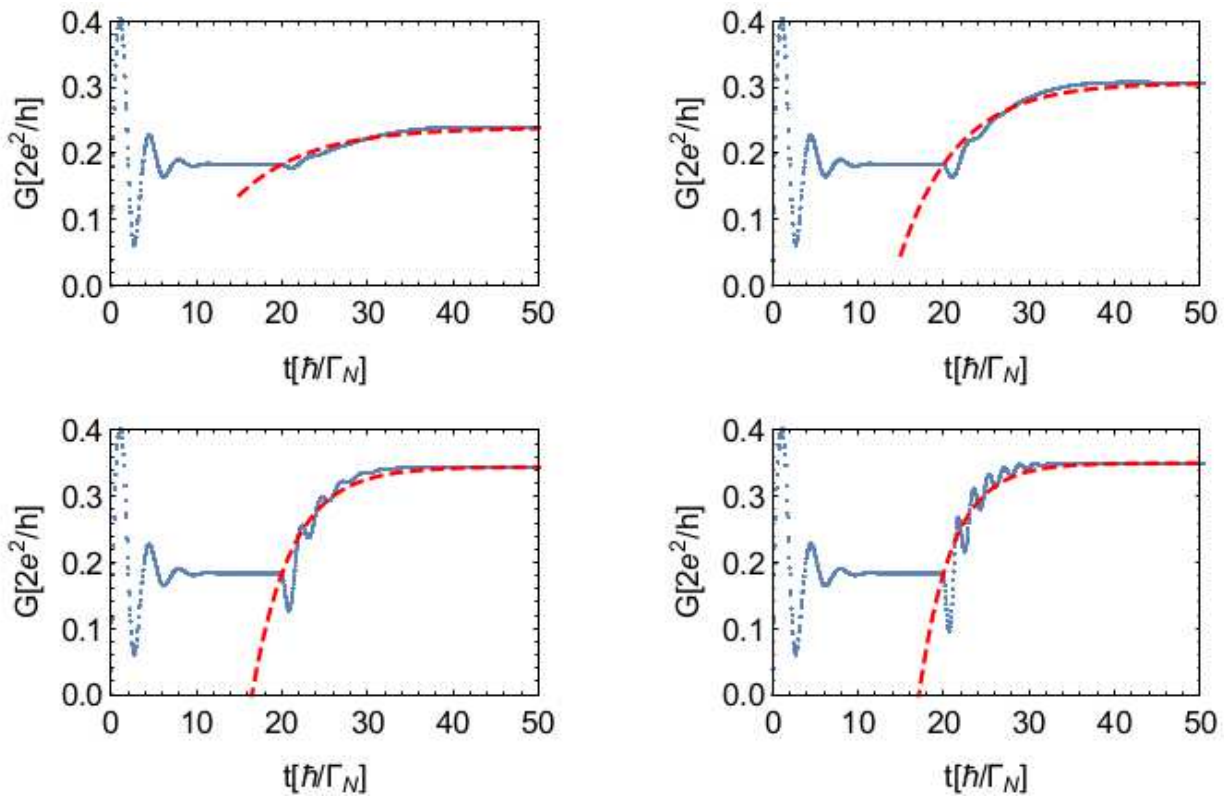}
	\caption{Blue dots represent time evolution of zero-bias differential conductivity obtained for $\Gamma_S=3\Gamma_N$, energy of the quantum dot $\epsilon_{\sigma}=\Gamma_{N}$ and quantum dot - chain couplings $t_m = 0.25\Gamma_{N}$ (upper left), $0.5\Gamma_{N}$ (upper right), $1\Gamma_{N}$ (lower left), $1.5\Gamma_{N}$ (lower right). Quantum dot is abruptly connected to Majorana mode at time $t=20\hbar/\Gamma_{N}$. Red dashed lines represent exponential fitting function with the  characteristic time scale $\tau$.} 
	\label{fitgrid}
\end{figure*}

Fig.~\ref{GridTemp2} shows the time-resolved differential conductance obtained for several values of the dot-chain hybridization $t_m$. We clearly notice, that the zero-energy feature develops more rapidly for stronger couplings $t_m$. Besides this zero-energy mode one also observes that the Andreev quasipartice states split into two branches. One of them (the low-energy Andreev branch) is located at $\pm \sqrt{\Gamma_S^2+\epsilon^2}$. The other (high-energy branch) arising from hybridization with the Majorana mode is formed  at $\pm \sqrt{\Gamma_{S}^2+\epsilon^2+2t_m^2}$, in agreement with the static solution predictions \cite{JBTDAK2017,JBTD2019}.

To quantify the time interval needed for formation of the zero-energy Majorana feature we have fitted (Fig.\ \ref{fitgrid}) a difference between the initial and final ($t=\infty$) zero-bias conductance by an exponential function
\begin{equation}
G(0,t) = G(0,\infty) - \left[ G(0,\infty) - G(0,t_{1})\right] e^{-(t-t_1)/\tau},
\label{eq:fit}
\end{equation}
where the initial moment $t_1=20$.
The phenomenological parameter $\tau$ characterizes the temporal interval, in which a mismatch between the initial conductivity and the equilibrium conductance diminishes $e$-times. Values of such numerically evaluated parameter  $\tau$ indicate, that development of the MZM features occurs the faster the stronger $t_{m}$ is. A few examples are listed in Tab.\ \ref{tab1}.

\begin{table}[h!]
  \begin{center}
    \begin{tabular}{c|c}
      \textbf{$t_m[\Gamma_{N}]$} & \textbf{$\tau [\hbar/\Gamma_{N}]$} \\
      \hline
 0.25 & 8.2 \\
 0.5 & 6.7 \\
 0.75 & 5.4 \\
 1 & 4.6 \\
 1.5 & 3.8 \\
    \end{tabular}
    \caption{Leakage time $\tau$ obtained for several couplings $t_m$.}
    \label{tab1}
  \end{center}
\end{table}

\section{Summary}
\label{sec:summary}

We have investigated time-resolved development of the Majorana features transmitted onto the quantum dot due to its coupling to the topological superconductor. For its feasible detection we have considered the tunneling of charge through a circuit, in which the quantum dot is strongly hybridized with the bulk superconductor and weakly coupled to the normal metallic lead. Our hybrid structure can be realized in practice, depositing the topological nanowire (for instance a chain of magnetic $Fe$ atoms) with the side-attached  quantum dot (e.g.\ nonmagnetic atom) on a surface of conventional superconductor \cite{Wiesendanger-2020,Wiesendanger-2018}. Approaching the conducting STM tip to QD its low-energy quasiparticles could be observable in the differential conductance, originating from the Andreev (particle-to-hole) scattering that is efficient in the low-bias regime, smaller or comparable to the superconducting gap.   
   
We have evaluated the characteristic time, needed for inducing the Majorana features in the zero-bias conductance. When the QD energy level is distant from the chemical potential, the Majorana leakage is manifested by enhancement of the differential conductance. In the opposite limit, when the quantum dot level is nearby the chemical potential, the Majorana mode has detrimental influence on the subgap spectrum of QD, producing the interferometric dip structure. We have evaluated the time interval during which these features emerge after: (i)  abrupt coupling of the quantum dot to the topological superconducting nanowire, and (ii) sudden change of the quantum dot energy level by external gate potential. 
For empirically realistic parameters  we have found, that emergence of the Majorana features would take in both cases about $2 - 20$ nanoseconds, in perfect agreement with estimations obtained by a full counting statistics analysis \cite{Souto-2020}. This dynamical process should be detectable with use of the currently available state-of-the-art tunneling spectroscopies. Such leakage seems to be fast enough to guarantee the practical realizations of braiding protocols designed for the Majorana  quasiparticles. 

In future studies it would be worthwhile to consider the correlated quantum dot, where the superconducting proximity effect competes with the Coulomb repulsion qualitatively affecting the in-gap bound states. In particular, they may cross one another at, so called, zero-pi transition.  Their dynamical interplay with the Majorana mode would require more sophisticated many-body techniques (e.g.\ time-dependent numerical renormalization approach) what is beyond a scope of our study.

\begin{acknowledgments}
This research has been conducted within a framework of the project {\em Analysis of nanoscopic systems coupled with superconductors in the context of  quantum information processing} No.\ GB/5/2018/209/2018/DA funded in the years 2018-2021 by the Ministry of National Defence Republic of Poland (JB, MB, TZ). 
The work is also supported by the National Science Centre (Poland) under the grant 2017/27/B/ST3/01911 (RT, TD).
\end{acknowledgments}

\begin{widetext}
\appendix*
\section{Equations of motion}
\label{sec:App}

In this appendix we present the explicit equations of motion for the momentum-dependent $f_{k i}$ functions 
\begin{eqnarray}
\frac{d f_{k 1}} {dt} &=& \left (i \epsilon_{\uparrow} - \frac{\Gamma_N}{2} \right ) f_{k 1} + i \frac{\Gamma_S}{2} f_{k 11} + i t_m (f_{k 5} + f_{k 6}) + i V_{k} e^{i\xi_{k}t}f_{FD}[(\xi_{k}-eV),T] \nonumber \\
\frac{d f_{k 2}} {dt} &=& \left (- i \epsilon_{\uparrow} - \frac{\Gamma_N}{2} \right ) f_{k 2} - i \frac{\Gamma_S}{2} f_{k 4} - i t_m (f_{k 7} + f_{k 8}) \nonumber \\
\frac{d f_{k 3}} {dt} &=& \left (- i \epsilon_{\uparrow} - \frac{\Gamma_N}{2} \right ) f_{k 3} - i \frac{\Gamma_S}{2} f_{k 12} - i t_m (f_{k 5} + f_{k 6}) \nonumber \\
\frac{d f_{k 4}} {dt} &=& \left (i \epsilon_{\downarrow} - \frac{\Gamma_N}{2} \right ) f_{k 4} - i \frac{\Gamma_S}{2} f_{k 2}
+ i V_{k} e^{i\xi_{k}t}f_{FD}[(\xi_{k}-eV),T] \nonumber \\
\frac{d f_{k 5}} {dt} &=& i \epsilon_{m} f_{k 5} + i t_m (f_{k 1} - f_{k 3}) \nonumber \\
\frac{d f_{k 6}} {dt} &=& - i \epsilon_{m} f_{k 6} + i t_m (f_{k 1} - f_{k 3}) \nonumber \\
\frac{d f_{k 7}} {dt} &=& - i \epsilon_{m} f_{k 7} + i t_m (f_{k 10} - f_{k 2}) \nonumber \\
\frac{d f_{k 8}} {dt} &=& i \epsilon_{m} f_{k 8} + i t_m (f_{k 10} - f_{k 2}) \nonumber \\
\frac{d f_{k 9}} {dt} &=& \left (- i \epsilon_{\downarrow} - \frac{\Gamma_N}{2} \right ) f_{k 9} + i \frac{\Gamma_S}{2} f_{k 10} \nonumber \\
\frac{d f_{k 10}} {dt} &=& \left (i \epsilon_{\uparrow} - \frac{\Gamma_N}{2} \right ) f_{k 10} + i \frac{\Gamma_S}{2} f_{k 9} + i t_m (f_{k 7} + f_{k 8}) \nonumber \\
\frac{d f_{k 11}} {dt} &=& \left (- i \epsilon_{\downarrow} - \frac{\Gamma_N}{2} \right ) f_{k 11} + i \frac{\Gamma_S}{2} f_{k 1} \nonumber \\
\frac{d f_{k 12}} {dt} &=& \left (i \epsilon_{\downarrow} - \frac{\Gamma_N}{2} \right ) f_{k 12} - i \frac{\Gamma_S}{2} f_{k 3}
\label{diff k functions}
\end{eqnarray}
and for the momentum-independent $f_i$ functions
\begin{eqnarray}
\frac{d f_1} {dt} &=& 2 \Im \left [- \frac{\Gamma_S}{2} f_4 + t_m (f_5 + f_6) + S_{k1} - \frac{i \Gamma_N}{2}  f_1 \right ] \nonumber \\
\frac{d f_2} {dt} &=& 2 \Im \left [- \frac{\Gamma_S}{2} f_4 + S_{k4} - \frac{i \Gamma_N}{2}  f_2 \right ] \nonumber \\
\frac{d f_3} {dt} &=& 2 \Im \left [t_m (f_6 - f_5) \right ] \nonumber \\
\frac{d f_4} {dt} &=& [- i (\epsilon_{\uparrow} +\epsilon_{\downarrow} ) - \Gamma_N] f_4 - i \frac{\Gamma_S}{2} (1 - f_1 - f_2) - i t_m (f_7 + f_8) + i (S_{k2} - S_{k 11}) \nonumber \\
\frac{d f_5} {dt} &=& \left [i (\epsilon_{\uparrow} -\epsilon_{m}) - \frac{\Gamma_N}{2} \right ] f_5 + i \frac{\Gamma_S}{2} f_7 + i t_m (f_3 - f_1) + i S_{k 5}^{*} \nonumber \\
\frac{d f_6} {dt} &=& \left [i (\epsilon_{\uparrow} +\epsilon_{m}) - \frac{\Gamma_N}{2} \right ] f_6 + i \frac{\Gamma_S}{2} f_8 + i t_m (1 - f_1 - f_3) + i S_{k 6}^{*} \nonumber \\
\frac{d f_7} {dt} &=& \left [-i (\epsilon_{\downarrow} +\epsilon_{m}) - \frac{\Gamma_N}{2} \right ] f_7 + i \frac{\Gamma_S}{2} f_5 - i t_m (f_4 + f_9) + i S_{k 7} \nonumber \\
\frac{d f_8} {dt} &=& \left [-i (\epsilon_{\downarrow} -\epsilon_{m}) - \frac{\Gamma_N}{2} \right ] f_8 + i \frac{\Gamma_S}{2} f_6 - i t_m (f_4 + f_9) + i S_{k 8} \nonumber \\
\frac{d f_9} {dt} &=& \left [-i (\epsilon_{\downarrow} -\epsilon_{\uparrow}) - {\Gamma_N}\right ] f_9 - i t_m (f_7 + f_8) + i (S_{k 12}^{*} - S_{k 10}) ,
\label{diff nok functions}
\end{eqnarray}
where $S_{k i}=\sum_{k} V_{k} e^{-i\xi_{k}t} \; f_{k i}$.

\end{widetext}

\bibliography{dynamics}

\end{document}